\documentclass[12pt]{article}
\usepackage{a4wide}
\usepackage{graphicx}
\DeclareMathAlphabet{\mathbbm}{U}{bbm}{m}{n}
  \SetMathAlphabet\mathbbm{bold}{U}{bbm}{bx}{n}
\begin{document}
\def\d{{\rm d}}
\title{\bf Baryon resonances in large $N_c$ QCD}
\author{N. Matagne
\thanks{{\it e-mail}: nmatagne@ulg.ac.be}~
and Fl.\ Stancu
\thanks{{\it e-mail}: fstancu@ulg.ac.be}
\\
{\small Institute of Physics, B5, University of Li\`ege,}\\[-6pt]
{\small Sart Tilman, 4000 Li\`ege 1, Belgium}%
}
\maketitle
%%%%%%%%%%%%%
\begin{abstract}\noindent
The  baryon spectra are discussed in the context of 
the $1/N_c$ expansion approach, with emphasis on
mixed symmetric states. The contributions of the spin dependent terms
as a function of the excitation energy are shown explicitly. At large
energies these contributions are expected to vanish.
\end{abstract}

\section{Introduction}
At energies corresponding to a length scale of the order of the hadron size
the standard perturbative QCD cannot be applied, because the coupling constant 
is too large. In the nonperturbative regime one can use the so-called
$1/N_c$ expansion approach which is based on the 32 years old 
idea of 't Hooft  \cite{HOOFT}, who suggested a perturbative expansion
of QCD in terms of the parameter $1/N_c$  where $N_c$ is the number
of colors. The double line diagrammatic method proposed by 
't Hooft  has been implemented 
by Witten  \cite{WITTEN} to describe hadrons by using
convenient power counting rules for Feynman diagrams.
According to Witten's intuitive picture, a baryon containing $N_c$ quarks 
is seen as a bound 
state in an average self-consistent potential of a Hartree type 
and the corrections to the Hartree approximation are of order $1/N_c$
which means that in the  $N_c \to \infty $ limit
the Hartree approximation is exact. Ground state baryons correspond to 
the ground state of the average potential. 

Ten years after 't Hooft's work, Gervais and Sakita  \cite{Gervais:1983wq}
and independently Dashen and Manohar in 1993  \cite{DM} discovered that QCD 
has an exact contracted SU(2$N_f$)$_c$ symmetry  
when $N_c \rightarrow \infty $,   $N_f$ being the number of flavors.
The contracted algebra generators acting in the spin-flavour space 
$X^{ia}$ are related to the SU(2$N_f$) generators  $G^{ia}$ 
in the limit $N_c \rightarrow \infty$ by
\begin{equation}
 X^{ia}=\lim_{N_c \to \infty} \frac{G^{ia}}{N_c}.
\end{equation}
 For ground state baryons the SU(2$N_f$) symmetry is broken by 
corrections proportional to $1/N_c$. Applications to ground state QCD 
baryons ($N_c$ = 3) were considered since 1993-1994.
Presently the $1/N_c$ expansion provides a systematic method to  
analyze baryon properties such as masses, magnetic moments, axial currents, etc.

The $1/N_c$ expansion method has been extended to excited states
since 1997 in the spirit of the Hartree approximation 
developed by Witten  \cite{Goity:1996hk}. It was shown that the SU($2N_f$) 
breaking occurs at order $N^0_c$, at variance with the ground state.
This conflict generated a conceptual problem, presently under investigation.

Here we are concerned with baryon spectra only.
If the SU($N_f$) symmetry is exact, the baryon mass operator is 
a linear combination of terms 
\begin{equation}
\label{massoperator}
M  = \sum_{i} c_i O_i,
\end{equation} 
with the operators $O_i$  having the general form
\begin{equation}\label{OLFS}
O_i = \frac{1}{N^{n-1}_c} O^{(k)}_{\ell} \cdot O^{(k)}_{SF},
\end{equation}
where  $O^{(k)}_{\ell}$ is a $k$-rank tensor in SO(3) and  $O^{(k)}_{SF}$
a $k$-rank tensor in SU(2), but invariant in SU($N_f$).
The latter is expressed in terms of  SU($N_f$) generators. 
For the ground state one has $k$ = 0. The first factor gives the 
order $\mathcal{O}(1/N_c)$ of the operator in the series expansion
and reflects Witten's power counting rules.
The lower index $i$ represents a specific combination of generators, see
examples below. 
In the linear combination, Eq. (\ref{massoperator}), each term of type (\ref{OLFS}) is multiplied 
by an unknown coefficient $c_i$ which is a reduced matrix element.
All these coefficients encode the QCD dynamics and are obtained from a fit to
the existing data. It is important to find regularities in their behaviour,
as shown below.
Additional terms are needed if SU($N_f$) is broken, as it is the case 
for $N_f = 3$ \cite{Matagne:2006zf}.

%%%%%%%%%%%%%%%%%%%%%%%%%%%%%%%%%%%%%%%%%%%%%%%%%%%%%%%%%%%%%%%%%%%%%%%%%%%
\section{The ground state}\label{se:gs}
A considerable amount of work has been devoted to ground state baryons
summarized in  several review  papers as, for example,
\cite{Dashen:1994qi,Jenkins:1998wy,Jenkins:2001it}.
The ground state is
described by the symmetric representation $[N_c]$. For $N_c$ = 3 
this becomes $[3]$ or  $[\bf 56]$
in an SU(6) dimensional notation. Let us consider below 
the simple case of two flavours, \emph{i.e.} SU(4). Its algebra is
\begin{eqnarray}\label{ALGEBRASU4}
&[S_i,S_j]  =  i \varepsilon_{ijk} S_k,
~~~~~ [T_a,T_b]  =  i \varepsilon_{abc} T_c,\nonumber \\
&[G_{ia},G_{jb}] = \frac{i}{4} \delta_{ij} \varepsilon_{abc} T_c
+\frac{i}{2} \delta_{ab}\varepsilon_{ijk}S_k.
\end{eqnarray}
As SU(4) is a group of rank 3 it has three invariants: $S^2, I^2$ and $G^2$.
\emph{i.e.} three operators of  type (\ref{OLFS}). But for the ground state 
one can take $I^2 = S^2$ in SU(4). Moreover due to the operator
identity \cite{Jenkins:1998wy}
\begin{equation}
 \{J^i,J^i\} + \{I^a,I^a\} + 4  \{G^{ia},G^{ia}\} = \frac{3}{2} N_c (N_c + 4)
\end{equation}
the invariant $G^2$ can be expressed in terms of $S^2$ and $I^2$.
So, one is left with one linearly independent operator, which we choose to be  
$S^2$. Accordingly, the mass formula takes the following simple form
\begin{equation}
M = m_0N_c+m_2\frac{1}{N_c}S^2+m_4\frac{1}{N_c^3}(S^2)^2
+\ldots+m_{N_c-1}\frac{1}{N_c^{N_c-2}}(S^2)^{N_c-3}.
\end{equation}
This describes a tower of large $N_c$ baryon states with 
$S = 1/2,...,N_c/2$, which collapses into a degenerate state 
when $N_c \to \infty$. One can see that the splitting starts at order $1/N_c$
when SU($2N_f$) is broken. 
The coefficients $c_i = m_i$ must be fitted from the data.
%%%%%%%%%%%%%%%%%%%%%%%%%%%%%%%%%%%%%%%%%%%%%%%%%%%%%%%%%%%%%%%%%%%%

\section{The excited states}\label{se:excit}

One expects 't Hooft's suggestion \cite{HOOFT} to hold
in all QCD regimes. Accordingly, the applicability of the $1/N_c$ 
expansion method to excited states is a subject of
current investigation. The  experimental facts indicate a small breaking
of SU(6) which make the $1/N_c$ studies of excited states entirely plausible.
%Presently the excited states are an open problem which is shortly
%reviewed here. Here the discussion is restricted to baryon masses.
The general form of a mass operator is given by Eq. (\ref{massoperator})  
%\begin{equation}
%\label{massoperator}
%M  = \sum_{i} c_i O_i.
%\end{equation} 
with $O_i$ defined as in Eq. (\ref{OLFS}).
For simplicity, here we discuss nonstrange baryons where SU(4) symmetry is exact.

Excited baryons can be divided into SU(6) multiplets, as in the
constituent quark model.
If an excited baryon belongs to the $[\bf{56}]$-plet
the mass problem can be treated similarly to the ground state
in the flavour-spin degrees of freedom, but one has to take into
account the presence of an orbital excitation in the space
part of the wave function  \cite{Goity:2003ab,Matagne:2004pm}.
If the baryon belongs to 
the mixed symmetric representation $[21]$, or $[\bf{70}]$ in SU(6)
notation, the treatment becomes much more complicated. In particular, the 
resonances up to 2 GeV belong to  $[{\bf 70},1^-]$, $[{\bf 70},0^+]$ or 
$[{\bf 70},2^+]$ multiplets.  

There is one standard way to study the   $[\bf{70}]$-plets
which is related to the Hartree approximation  \cite{Goity:1996hk}.
This consists in reducing the description of an excited baryon  to that
of an excited quark coupled to a symmetric core, see \emph{e.g.} 
 \cite{Carlson:1998vx,Goity:2002pu,Matagne:2005gd,Matagne:2006zf}.
 In that case the core 
can be treated in a way similar to that of the ground state.
In this method each SU($2N_f$) $\times$ O(3) generator is  splitted 
into two terms 
\begin{equation}\label{CORE}
S^i = s^i + S^i_c; ~~~~T^a = t^a + T^a_c; ~~~ G^{ia} = g^{ia} + G^{ia}_c,
~~~ \ell^i = \ell^i_q + \ell^i_c,
\end{equation}
%and for the SO(3) generators $\ell^i$ one has 
%\begin{equation}\label{SO3}
%\ell^i = \ell^i_q + \ell^i_c,
%\end{equation}
where  $s^i$, $t^a$, $g^{ia}$ and $\ell^i_q$  are the excited 
quark operators and  
$S^i_c$, $T^a_c$, $G^{ia}_c$ and  $\ell^i_c$ the corresponding core operators.
As an example, we discuss the latest in date results, for  nonstrange 
baryons belonging to the $[{\bf 70},\ell^+]$ multiplets with $\ell$ = 0 and 2.
The list of the dominant operators up to order $1/N_c$ is given in Table 1
together with the values of the coefficients 
$c_i$ obtained from the data. It is customary to drop corrections of order
$1/N^2_c$.
In this list, the first is the trivial
operator of order $\mathcal{O}(N_c)$. The second is the 1-body part of the spin-orbit 
operator of order $\mathcal{O}(1)$ which acts on the excited quark.
The third is a composite 2-body operator formally of order $\mathcal{O}(1)$ as 
well. It involves 
the tensor operator %(\ref{TENSOR}) 
\begin{equation}\label{TENSOR}
\ell^{(2)ij}_{q}=\frac{1}{2}\left\{\ell^i_q,\ell^j_q\right\}
-\frac{1}{3}\delta_{i,-j}\vec{\ell}_q\cdot\vec{\ell}_q~,
\end{equation}
acting on the excited quark and the SU(6)
generators $g^{ia}$ acting on the excited quark and $G^{ja}_c$ acting on the
core. The latter is a coherent operator which introduces an extra power $N_c$
so that the order of $O_3$ is $\mathcal{O}(1)$.

\begin{table}[tbp]
%\vspace{2cm}
\begin{center}
\renewcommand{\arraystretch}{1.25}
\begin{tabular}{llrrl}
\hline
\hline
Operator & \multicolumn{4}{c}{Fitted coef. (MeV)}\\
\hline
\hline
$O_1 = N_c \mathbbm{1}  $                           & \ \ \ $c_1 =  $  & 555 & $\pm$ & 11       \\
$O_2 = \ell_q^i s^i$                                & \ \ \ $c_2 =  $  &   47 & $\pm$ & 100    \\
$O_3 = \frac{3}{N_c}\ell^{(2)ij}_{q}g^{ia}G_c^{ja}$ & \ \ \ $c_3 =  $   & -191 & $\pm$ & 132  \\
$O_4 = \frac{1}{N_c}(S_c^iS_c^i+s^iS_c^i)$          & \ \ \ $c_4 =  $  &  261 & $\pm$ &  47  \\
\hline \hline
\end{tabular}
\caption{List of operators and the coefficients resulting from the fit 
of nonstrange baryon masses assumed to belong
to the $[{\bf 70},0^+]$- and $[{\bf 70},2^+]$-plets. The  fit gave
 $\chi^2_{\rm dof}  \simeq 0.83$ \ \cite{Matagne:2005gd}.}
\label{operators}
\end{center}
\end{table}

In this procedure, there are two major drawbacks, related to each other.
The first is that the number of  linearly
independent operators constructed from the generators given in the
right-hand side of Eqs. (\ref{CORE}) 
increases tremendously so that the number of coefficients 
to be determined becomes much larger than the experimental data
available. Consequently, in selecting the most dominant operators 
one has to make an arbitrary choice, as we did in Table  \ref{operators},
similarly to previous literature.
\begin{figure}\label{COEF}
\begin{center}
\includegraphics[width=10cm]{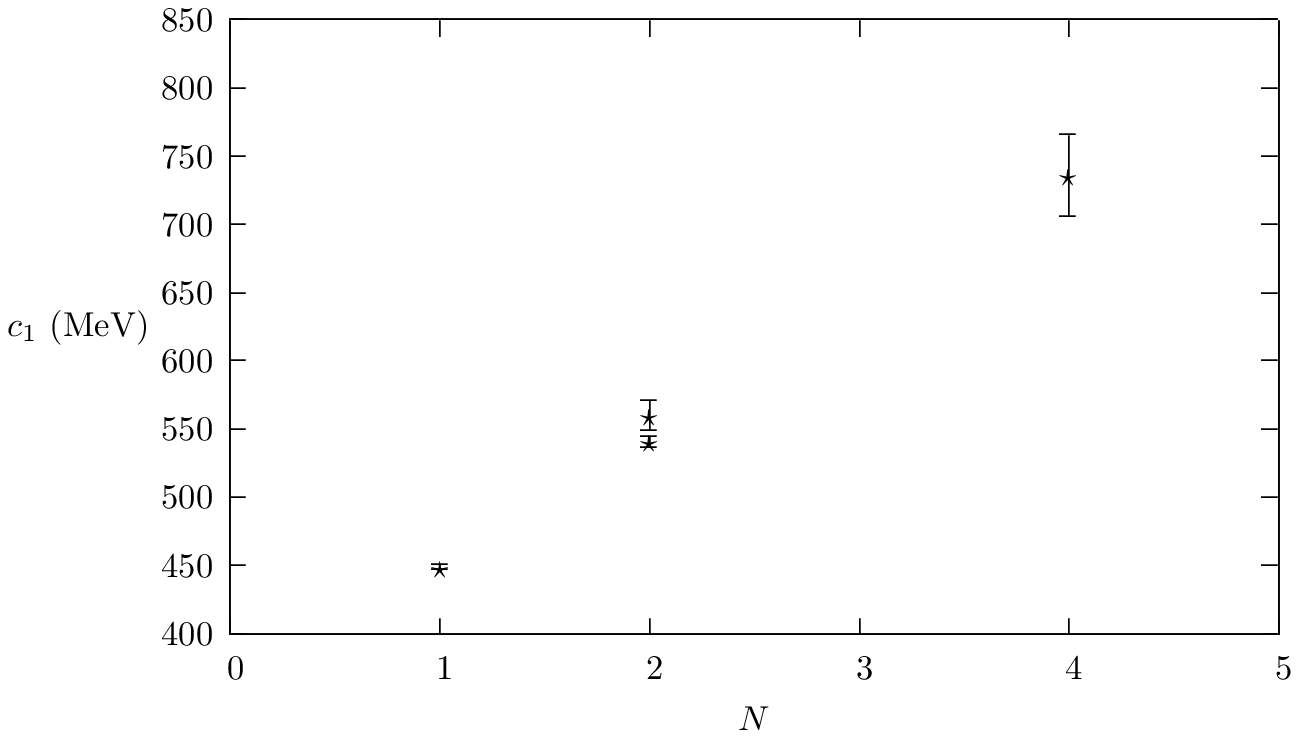} \\
\vspace{0.5cm}
\includegraphics[width=10cm]{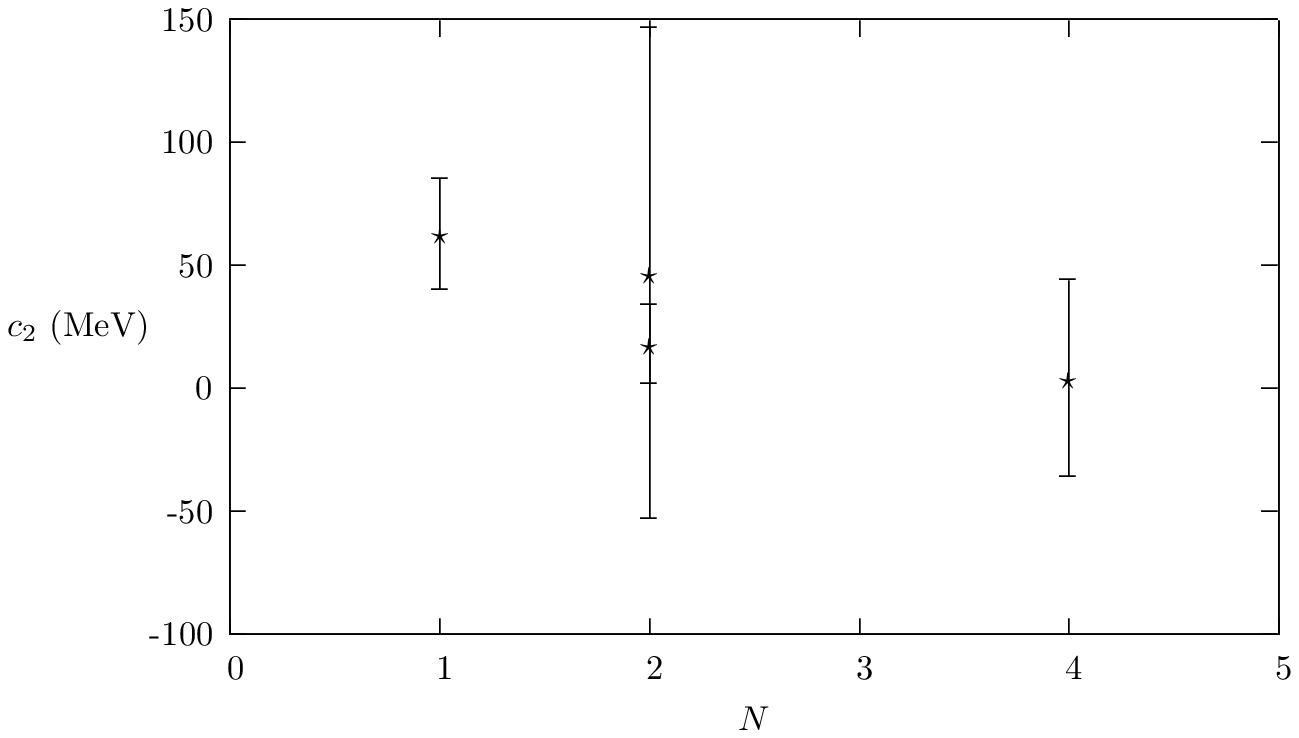} \\
\vspace{0.5cm}
\includegraphics[width=10cm]{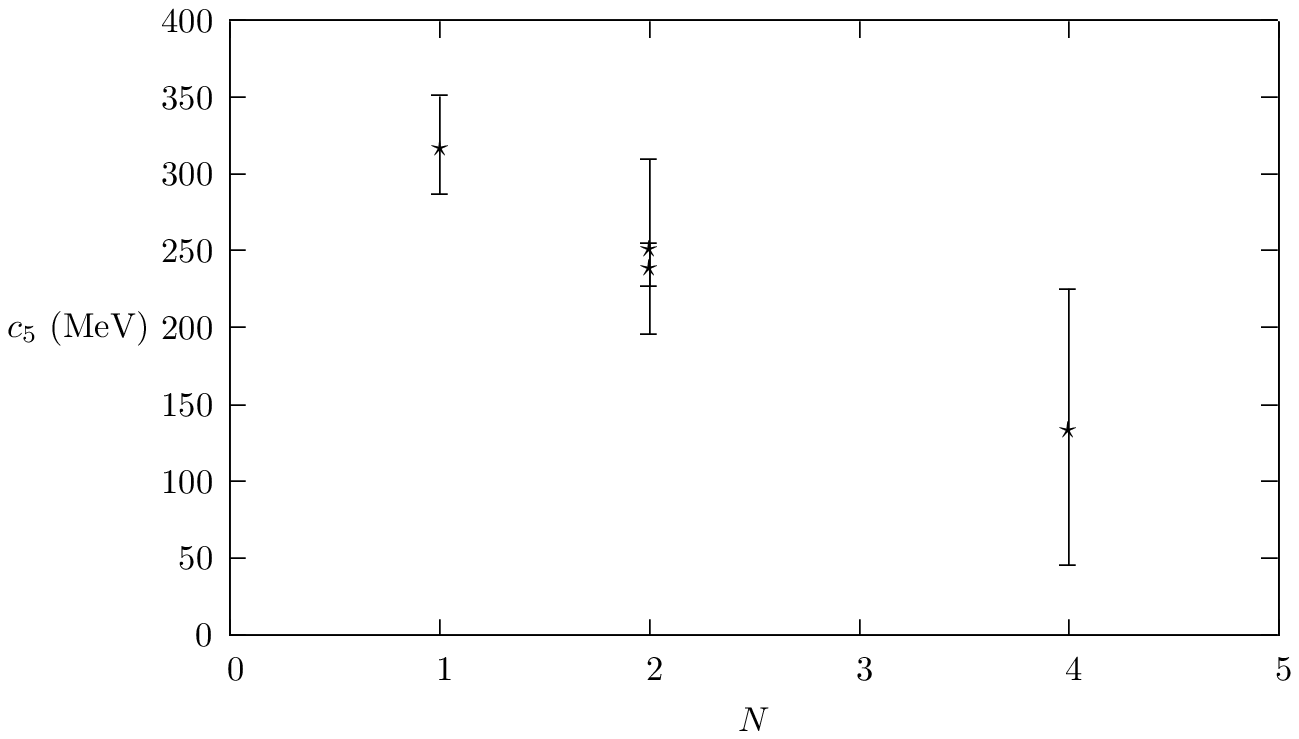}
\vspace{-0.2cm}
\caption{The coefficients $c_i$ vs $N$ from various sources:
for $N = 1$ from Ref. \cite{Goity:2002pu}
for $N = 2$ from Refs. \cite{Goity:2003ab} (lower values) and 
\cite{Matagne:2005gd} (upper values),
for $N = 4$ from Ref. \cite{Matagne:2004pm}.} 
%The straight lines are to guide the eye.
\end{center}
\end{figure}
The second drawback is  related to the truncation of the available basis 
vector space. Among the basis vectors of
the irreducible representation  $[N_c-1,1]$ of $S_{N_c}$, 
in this procedure only the vector corresponding to the normal 
Young tableau is kept, the reason being to decouple the system into a symmetric 
core and an excited quark. 
In a normal Young tableau this can be easily done by removing 
the $N_c$-th particle from the second row. The terms represented 
by the other possible 
Young tableaux, needed to construct a symmetric orbital-flavour-spin
state are neglected, \emph{i.e.} antisymmetry is ignored. 
As a result the procedure brings in terms of order $N^0_c$, which is 
in conflict with the $1/N_c$ expansion for the ground state.  

A solution to this problem has been found in Ref.  \cite{Matagne:2006wi},
where the separation into a symmetric core and an excited quark is
avoided through the calculation of the matrix elements of
the SU(4) generators by using  a generalized Wigner-Eckart 
theorem \cite{Hecht:1969ck}.
In this way the antisymmetry is properly taken into account.
% and previous analytic results \cite{Hecht:1969ck} derived in the 
%context of nuclear physics.  
The result is that the $1/N_c$ expansion
starts at order $1/N_c$, as for the ground state.

Based on group theory  arguments it is expected that the mass 
  splitting starts at order $1/N_c$, as a general rule, 
  irrespective of the angular momentum and parity of the state and also 
  of the number of flavours, provided SU($N_f$) is an exact 
  symmetry. 
  
Despite the drawbacks of the splitting method, the application 
of the   $1/N_c$ expansion method gave useful results, as a first
approximation. They predicted the behaviour of the coefficients $c_i$ in the
mass formula as a function of the excitation energy. This 
is illustrated in Fig. 1. This figure suggests that the spin-orbit and
the spin-spin terms vanish with the excitation energy, bringing a 
strong support to constituent quark models and that the spin-spin term
is dominant among all the other spin dependent terms.
Note that in a quark model picture, the coefficient $c_1$
would correspond to the additional contribution of a free mass term, 
the kinetic energy and the confinement. It is not thus surprising that 
it raises with the excitation energy. 
	
%%%%%%%%%%%%%%%%%%%%%%%%%%%%%%%%%%%%%%%%%%%%%%%%%%%%%%%%%%%%%%%%%%%%%

\section{Conclusion}\label{se:concl}

The $1/N_c$ expansion method provides a powerful theoretical
tool to analyze the spin-flavour symmetry of baryons and
explains the succes of models based on spin-flavor symmetry.

%%%%%%%%%%%%%%%%%%%%%%%%%%%%%%%%%%%%%%%%%%%%%%%%%%%%%%%%%%%%%%%%%%%%%%%%%%%%%%%%
 
\end{document}